\documentclass[universe,article,accept,pdftex,moreauthors]{Definitions/mdpi} 

\firstpage{1} 
\pubvolume{1}
\issuenum{1}
\articlenumber{0}
\pubyear{2024}
\copyrightyear{2024}
\datereceived{ } 
\daterevised{ } 
\dateaccepted{ } 
\datepublished{ } 
\hreflink{https://doi.org/} 

\Title{Cosmological test of an ultraviolet origin of Dark Energy}

\TitleCitation{UV origin of Dark Energy}

\Author{Hans Christiansen, Bence Tak\'acs  and Steen H. Hansen \ddag *}

\AuthorNames{Hans Christiansen, Bence Tak\'acs and Steen H. Hansen}

\AuthorCitation{Christiansen, H.; Tak\'acs, B.; Hansen, S. H.}

\address{%
Dark Cosmology Center, Niels Bohr Institute, Jagtvej 155, 2100 Copenhagen, Denmark }

\corres{Correspondence: hansen@nbi.ku.dk}

\secondnote{These authors contributed equally to this work.}

\abstract{The accelerated expansion of the Universe is impressively well described by a cosmological constant. However, the observed value of the cosmological constant is much smaller than expected based on quantum field theories.
Recent efforts to achieve consistency in these theories have proposed a relationship between Dark Energy and the most compact objects, such as black holes (BH). However, experimental tests are very challenging to devise and perform.
In this article, we present a testable model with no cosmological constant, in which the accelerated expansion can be driven by black holes. The model couples the expansion of the Universe (the Friedmann equation) with the mass-function of cosmological haloes (using the Press-Schechter formalism).
Through the observed link between halo-masses and  
BH-masses one thus
gets a coupling between the expansion rate of the Universe
and the BHs.
\color{black}
We compare the predictions of this simple BH model with SN1a data and find a poor agreement with observations.
Our method is sufficiently general that it allows
us to also test 
a fundamentally different model, also without
a cosmological constant, where
the accelerated expansion is driven by 
a new force proportional to the internal
velocity dispersion of galaxies. Surprisingly
enough this model cannot be excluded
using the SN1a data.
\color{black}
}


\begin{document}
\section{Introduction}
The accelerated expansion of the Universe was originally observed in SN1a data~\cite{1999ApJ...517..565P, 1998AJ....116.1009R}. Subsequently, these findings have been confirmed by a range of independent observations, including the growth of the large-scale structures and the cosmic microwave 
background~\cite{2011ApJS..192...18K, 2010MNRAS.401.2148P, 2011MNRAS.415.2876B, 2011ApJS..192...16L, 2009ApJ...700.1097H, 2011MNRAS.418.1725B}
all of which indicate that a cosmological constant, represented by $\Lambda$, apparently provides 
excellent agreement with all observables. This is quite remarkable because it implies that the cosmological standard model fits nearly all astronomical observations with just a handful of free parameters, one of which is the energy density represented by the cosmological constant $\Lambda \approx 3 \times 10^{-122} l_{\rm P}^{-2}$, where $l_{\rm P}$ is the Planck length.

However, a significant problem arises as a quantum field explanation of the magnitude of $\Lambda$ is off by approximately 120 orders of magnitude~\cite{1989RvMP...61....1W, 2003RvMP...75..559P}. This discrepancy has led theoretical physicists to contemplate: 
{\it "If a solution to the cosmological constant exists, it may involve some complicated interplay between infrared and ultraviolet effects (maybe in the context of quantum gravity)"}
\cite{2017arXiv171007663G}. 

The concept of linking the largest scales (cosmological constant on cosmological scales) with the most compact objects (such as black holes) was explored by Cohen et al~\cite{1999PhRvL..82.4971C}. They discussed effective field theories with a cut-off scale $\lambda$, where the entropy in a box of volume $L^3$ is $S \sim L^3 \lambda^3$. However, the Beckenstein entropy \cite{1973PhRvD...7.2333B, 1981PhRvD..23..287B}  of a black hole has a maximum value of $S_{Be} \sim L^2$. This discrepancy may lead to inconsistencies when dealing with very large objects like the entire Universe.
To address this issue, Cohen et al~\cite{1999PhRvL..82.4971C} proposed a relationship between the UV cut-off and the IR physics to ensure that effective field theories remain consistent. This idea has garnered significant interest in the theoretical physics community over the last few years 
\cite{2021PhRvD.104g6024B,2021PhRvD.104l6032A, 2020JHEP...03..037C}.

One crucial, missing element between the observation of the accelerated expansion of the Universe and the range of theories suggesting a connection between the IR and UV phenomena is a testable model. In this article, we present a phenomenological model that contains no cosmological constant. Instead, the model calculates the time-dependence of the Universe's expansion based on the 
evolution of the abundance of large-scale structures.
Cosmological structure formation follows a bottom-up process, where small structures merge to form larger structures, resulting in a time-evolution of this new effect.

Since it is uncertain whether the UV-IR connection should be fundamentally linked to the entropy of black holes raised to some power \cite{1999PhRvL..82.4971C, 2021PhRvD.104g6024B}, the velocity dispersion of dark matter in cosmological haloes \cite{2021ApJ...910...98L, 2021MNRAS.506L..16H}, or something entirely different, we introduce a single parameter, denoted as $\beta$, along with a normalization, to encompass all these cases. This way, we introduce a new "force" that is proportional to the sum of $\sum M_{\rm halo}^\beta$, where $M_{\rm halo}$ represents the mass of cosmological haloes.
By comparing the resulting cosmological expansion with SN1a data, we find that this phenomenological model, without a cosmological constant, provides a temporal evolution that appears to
be as approximately as good as the standard $\Lambda$CDM model.

\section{The basic idea}
The expansion of the Universe is independent of the
amount of haloes in the standard description of cosmology. 
This is seen by the fact
that the Friedmann equation, which describes the
expansion of the Universe, can be written as
\begin{equation}
    \left( \frac{H}{H_0} \right)^2 = \Omega_{\rm{M},0} \, a^{-3} + \Omega_{\Lambda,0} \, ,
\end{equation}
where the Hubble parameter is given by $H = \dot a/a$, $a(t)$ is
the radius of the Universe normalized to unity today, and
all quantities with sub-0 represent quantities today, 
such as $\Omega_{\rm{M},0} =0.3$
and $\Omega_{\Lambda,0}=0.7$. This equation may be described by
$a(t) = a(\Omega_M, \Omega_\Lambda)$, and one can include 
terms for radiation 
\color{black} and curvature \color{black}
in the equation as well. 

Knowing the expansion history of the Universe, one can now
calculate the number of haloes of a given mass as a function
of time $N(M, t)$. One example of this is given by the Press-Schechter
formalism \cite{1974ApJ...187..425P}, which will be 
discussed in detail below. Using the fact that
the expansion is a function of time $a(t)$,
the distribution of haloes can be described by $N(M, a(t))$.

Instead, as will be shown below, by introducing a new energy-term
related to the distribution of haloes, one can get a new
Friedmann equation, which looks like
\begin{equation}
    \left( \frac{H}{H_0} \right)^2 = \Omega_{\rm{M},0} \, a^{-3} \left( 
    1 + F\left[ N \left( M, a \right) \right] \right) \, ,
\end{equation}
with no cosmological constant.
The function $F[N]$ depends on the distribution of masses of cosmological
haloes. As a concrete example, one can use the observed connection between the halo masses and the black hole masses
\color{black}
(extrapolated to be valid at all masses),
\color{black}
one thus sees that the expansion may be written as a function
of the distribution of BH masses. The change from the standard Friedmann equation to this model can hence be described by
\begin{equation}
    a \left( \Omega_M, \Omega_\Lambda \right) \rightarrow a\left( \Omega_M, N(M,a) \right) \, .
\end{equation}

\color{black}
It is important 
to clarify the following point. Observational
data, such as that from CMB and SN1a,
show that Eq.~(1) provides an excellent
fit with an essentially flat Universe.
If we instead 
calculate the expansion of a
universe using Eq.~(2), then one
may get an accelerated expansion
very similar to that of the
$\Lambda$CDM model. This implies
that if we were to analyse the 
corresponding data in that universe 
from CMB or SN1a 
with Eq.~(1), then we would again 
conclude that the universe is flat.
A detailed discussion on this point
was made by Linder \& Jenkins
\cite{2003MNRAS.346..573L}
who wrote the corresponding
RHS of our Eq.~(2) as 
$\Omega_M a^{-3} + \delta H^2/H_0^2$, and
they wrote: 
"all we have observed for sure
is a certain energy density due to matter, $\Omega_m$, and consequences of the expansion rate $H(z)$".
\color{black}

\section{The Press-Schechter formalism}

The evolution of the number of structures of mass $M$ 
as a function of cosmic time was first derived by
Press \& Schechter~\cite{1974ApJ...187..425P}. 
Under the assumption that 
 primordial density perturbations are Gaussian, the distribution of the amplitudes of perturbations of mass $M$ will take the form
\begin{equation}
    p(\delta) = \frac{1}{2 \sqrt{\pi} \sigma(M)} \exp{\left[ -\frac{\delta^{2}}{2 \sigma^{2}(M)} \right]} \, ,
\end{equation}
where the density contrast of a perturbation of mass $M$ is defined as $\delta = \frac{\delta \rho}{\rho}$ and $\sigma(M)$ is the variance. 
Such a distribution will have its variance equal to the mean of the square of density fluctuations $\sigma^{2}(M) = \left< \delta^{2} \right>$. Press and Schechter assumed that 
 upon reaching some critical amplitude $\delta_{c}$, density perturbations will rapidly form into bound objects. 

The variance of density perturbations $\sigma^{2}(M)$ is directly related to the mass $M$ of bound density perturbations and to the power spectrum of density perturbations $P(k)$ by
\begin{equation} \label{eqn:sigma-mass}
    \sigma^{2}(M) \propto AM^{-(n+3)/3} \, ,
\end{equation}
where $n$ is the spectral index.
\color{black}
Throughout this paper we assume that $n\approx -2.5$ as observed at galaxy scales today 
\cite{2008cosm.book.....W, 2008gady.book.....B}.
\color{black}
The fraction $F(M)$ of fluctuations of masses within the range $M$ to $M + \mathrm{d}M$ which become bound at epoch $t_{c}$ for amplitudes $\delta > \delta_{c}$ is
\begin{equation} \label{eqn:mass-fraction}
    F(M) = \frac{1}{\sqrt{2} \pi \sigma(M)} \int_{\delta_{c}}^{\infty} \exp \left[ - \frac{\delta^{2}}{2\sigma^{2}(M)} \right] \mathrm{d}\delta  \, ,
\end{equation}
where $t_{c} = \delta_{c} / \sqrt{2}\sigma(M)$ is the critical time.
The critical time $t_{c}$ is related to the mass distribution $M$ by the relation (\ref{eqn:sigma-mass}) and is rewritten
\begin{equation}
    t_{c} = \frac{\delta_{c}}{\sqrt{2}\sigma(M)} = \left( \frac{M}{M^{\ast}} \right)^{(3+n)/6} \, ,
\end{equation}
where $M^{\ast} = (2A/\delta_{c}^{2})^{3/(3+n)}$ is a reference mass wherein information on cosmic epoch is contained.  
The fluctuations evolve according to $\ddot \delta + 2H \dot \delta = 4 \pi G \rho \delta$, and 
from \cite{1977MNRAS.179..351H, 1992ARA&A..30..499C} it is known that in homogeneous and isotropic cosmologies the amplitudes of density perturbations grow according to
\begin{equation}
    \delta(a) \propto \frac{\dot{a}}{a} \int_{0}^{a} \frac{\mathrm{d}a'}{(\dot{a}')^{3}} \, .
\end{equation}
This equation is valid even though the expansion history
is not given by a $\Lambda$CDM model, however, as
we will find that the expansion history is surprisingly
close to that of $\Lambda$CDM, the evolution of
$\delta (a)$ will be very close to that in a
$\Lambda$CDM universe.
We will, never the less, solve this equation
numerically as a function of the actual expansion
history of our model.

We can now incorporate time implicitly into $M^{\ast}$ as
\begin{equation} \label{eqn:ref-mass}
    M^{\ast} = M_{0}^{\ast} \left( \frac{\delta(a)}{\delta(a_{0})} \right)^{6 / (3 + n)} \, .
\end{equation}

By assuming that $M = \Bar{\rho} V$, $\Bar{\rho}$ is the mean density of the background, and $V$ is the volume, one obtains
\begin{equation} 
    N(M) = \frac{\Bar{\rho}}{\sqrt{\pi}} \frac{\gamma}{M^{2}} 
    \left( \frac{M}{M^{\ast}} \right)^{\gamma/2} 
    \exp \left[ -\left(\frac{M}{M^{\ast}}\right) ^{\gamma} \right] \, ,
    \label{eq:massfunction}
\end{equation}
where $\gamma = 1 + \frac{n}{3}$.
The above derivation is standard and can be found in
many textbooks \cite{2008gafo.book.....L}.

With this expression we can now calculate 
the expectation value of a power of $M$  as
\begin{equation} \label{eqn:exp_mass_alpha}
    \left< M^{\alpha} \right> = \int_{0}^{\infty} N(M) M^{\alpha} \mathrm{d}M \, ,
\end{equation}
and if we have ratios of such expectation values, the
normalizations cancel
\begin{equation}\label{eqn:massmoments}
    \frac{\left< M_{i}^{\beta+1} \right>}{\left< M_{i} \right>} = \frac{\int_{0}^{\infty} M^{\beta+\gamma/2-1} \exp\left[-{(M/M^{\ast})^{\gamma}}\right] \mathrm{d}M}{\int_{0}^{\infty} M^{\gamma/2-1} \exp\left[{-(M/M^{\ast})^{\gamma}}\right] \mathrm{d}M} = {M^*}^\beta I\, ,
\end{equation}
and the integrals can be expressed through Gamma-functions.

\section{The revised Friedmann equation}

If one considers a new force proportional to the squared
velocity dispersion of the dark matter particles in a
cosmological halo, then this leads to an extra term
in the Friedmann equation \cite{2021ApJ...910...98L} 
\begin{equation}
    \left( \frac{H}{H_0} \right) ^2 = \Omega_{\rm{M},0} a^{-3}  \left[ 1 + \eta \sum \left( \frac{\sigma_i}{c} \right)^2 \right] \, ,
\end{equation}
and if one instead considers the change of energy to arise
from a more general term
\begin{equation}
    \Delta E = - \kappa \frac{Gm}{r} \sum_{i} M_{i}^{\beta + 1} \, ,
\end{equation}
one   gets a new Friedmann equation of the form
\begin{equation}
    \left( \frac{H}{H_0} \right) ^2 = \Omega_{\mathrm{M}, 0} a^{-3} \left( 1 + \kappa \frac{\left< M^{\beta + 1}_{i} \right>}{\left< M_{i} \right>} \right) \, .
\end{equation}
Defining the constant $\mu = \kappa I M^{* \beta}_0 \,\delta(a_0) ^{-2/\gamma}$ and using Eq. (\ref{eqn:massmoments}) one obtains
\begin{equation}
    \left( \frac{H}{H_0} \right) ^2 = \Omega_{\mathrm{M}, 0} a^{-3} \left( 1 + \mu f  \right) \, ,
\label{eq:newfriedmann}
\end{equation}
where

\begin{equation}
    f = \left( \frac{\dot{a}}{a} \int_{0}^{a} \frac{\mathrm{d}a'}{(\dot{a}')^{3}} \right)^{1/p} \, ,
\end{equation}
and $p=\gamma/2\beta$.

\color{black}
The effect of the terms on the RHS of
Eq.~(16) can be analysed just like
in the standard cosmology, where
each term can be described by an 
equation of state with 
properties $\rho_j = a^{-3(1+\omega_j)}$.
Thus the first term (which is just
the CDM) leads to $\omega_m = 0$, and 
the second term may lead to 
$\omega_f = -1$ in the case
that the choice of $\mu$ and $\beta$
happens to lead to an expansion history
similar to that of the $\Lambda$CDM
Universe, as we will show below may
happen for carefully chosen values.

Whereas the entire RHS of Eq.~(16)
may be viewed as resulting from 
CDM, then the difference in the
temporal evolution of the two
terms is crucial:
It has been observed that the
Universe transitions from 
a positive deceleration
parameter $q=-\ddot a a /\dot a^2$
to a negative one 
\cite{2004ApJ...607..665R}. 
In Eq.~(16) the corresponding
early effect is driven by the first 
term (the standard CDM term) and the transition
to the late accelerated evolution is driven by
the second $\mu f$-term.
\color{black}

We now have the new Friedmann equation~(\ref{eq:newfriedmann})
which must be solved numerically. Since this model
can mimic the accelerated expansion of the universe
through the evolution of all the overdensities, we
will below refer to this model as a $\delta$CDM model.
There are, in principle, two free parameters, namely
$\beta$ which should come from some fundamental principle
(as described in the introduction) and $\mu$ which 
is merely a normalization of this effect.
Equation (\ref{eq:newfriedmann}) 
contains the overdensities $\delta$
on the RHS
and is thus significantly more
complex than the Friedmann equation
of the $\Lambda$CDM model.
For the numerical solution we use a backwards differentiation formula, which is an implicit method of numerical integration suited to stiff problems.

\section{Supernova data}

In order to test the model we compare with SN1a data on the apparent magnitude.
In this work we employ SN1a data from the Pantheon+ analysis \cite{2022ApJ...938..113S}, which includes 1550 SN1a of redshifts up to $z \sim 2$.
We also calculate a simple $\chi^2$ to estimate the quality of the
models, as compared to the standard $\Lambda$CDM model,
and leave a proper analysis including the
covariance matrix, allowing $\Omega_M$ 
or the spectral index $n$ to be  
\color{black} scale- or time-dependent, 
\color{black} etc to the future.

\begin{figure}[hbt]
	\includegraphics[width=\columnwidth]{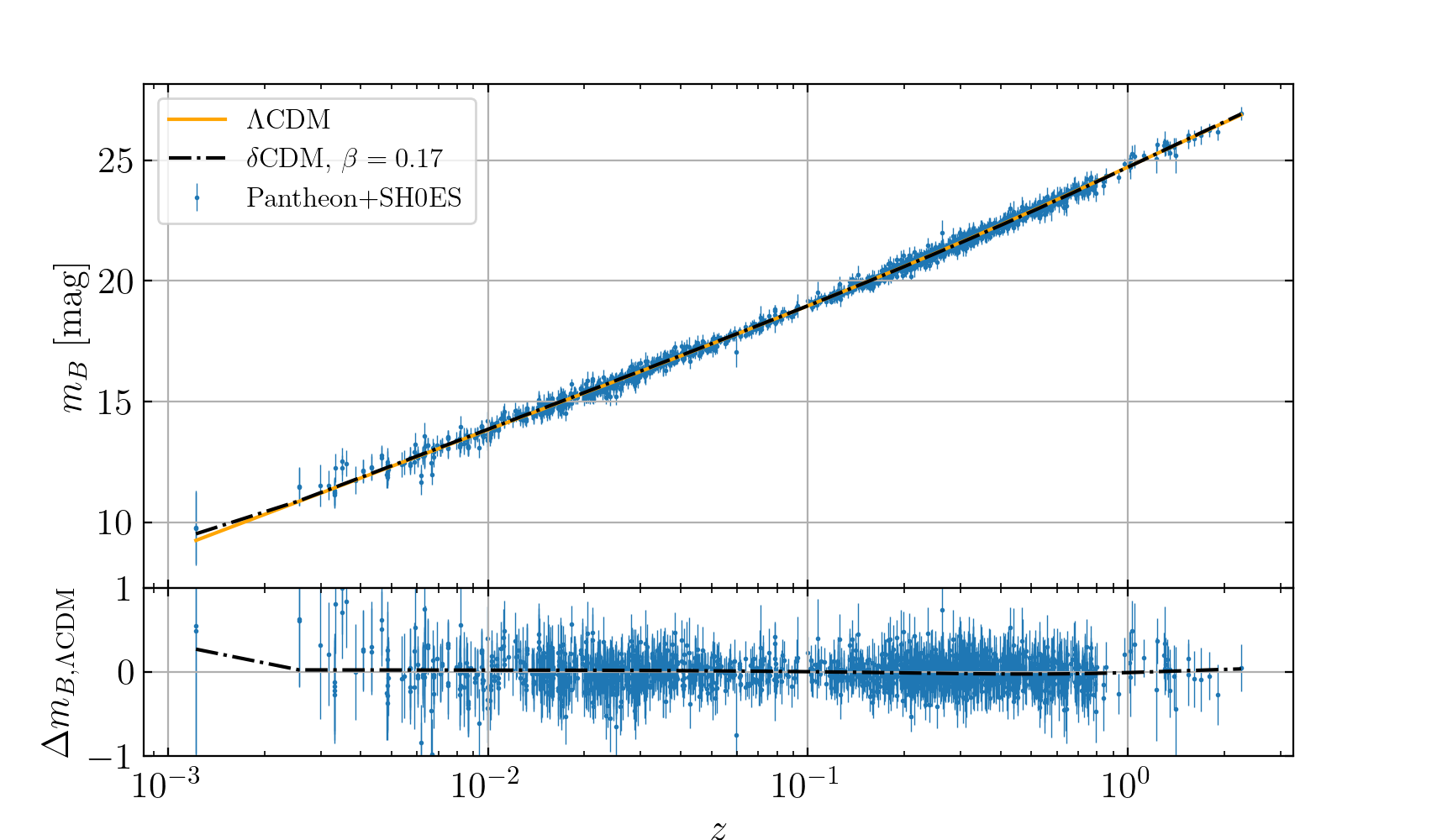}
    \caption{
  Apparent magnitudes of SN1a from the Pantheon+ analysis \cite{2022ApJ...938..113S} overlaid with the 
  apparent magnitudes of a flat-$\Lambda$CDM model and a $\delta$CDM model with
    $\beta = 0.17$. The 
    residuals from $\Lambda$CDM model 
        are shown below. The alternative model is seen to 
        follow the expansion history of the Universe in
        fairly good agreement with the SN1a data.  
        The small kink at $z\approx 10^{-3}$ is 
        due to the finite steps used in $z$.}
    \label{fig:sn}
\end{figure}

In Fig.~\ref{fig:sn} we present the SN1a data together with the standard $\Lambda$CDM model ($\Omega_{\rm{M},0}=0.33$, 
$\Omega_{\Lambda,0} = 0.67$), and a $\delta$CDM model with
$\beta = 0.17$ (and the best fit normalization $\mu$). 
It is clear that the two models approximately follow the SN1a
equally well. The $\chi^2$ of the $\delta$CDM is \color{black}
slightly
bigger \color{black} than that of the $\Lambda$CDM model. In the lower
panel we show the residual from the $\Lambda$CDM model.

\begin{figure}
	\includegraphics[width=\columnwidth]{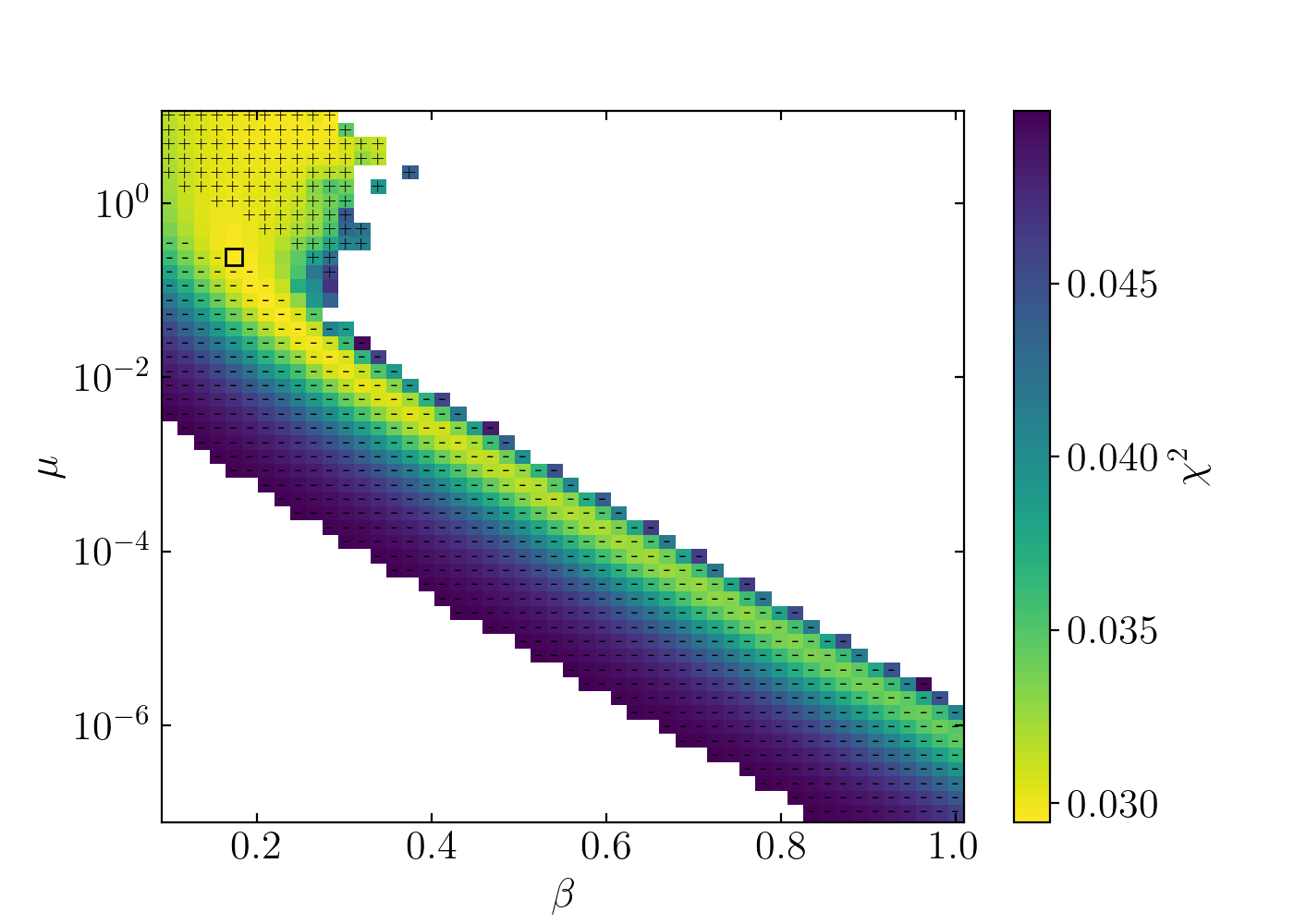}
    \caption{A grid in $\beta$ and $\mu$. The colour indicates
    $\chi^2$ calculated over the SN1a data. All points with
    $\chi^2$ significantly worse than that of $\Lambda$CDM
    have been left white. There are several points in the
    parameter space which have models that fit the SN data
    as approximately as well as $\Lambda$CDM, including points in the range
    $\beta =0.15-0.4$. 
All points with
absolute magnitude $M > -19.0$ are indicated with 
plus-sign, and a minus-sign for $M < -19.5$.  
    The black square at $\beta=0.17$ represent 
    the model of figure \ref{fig:sn}.}
    \label{fig:scan}
\end{figure}

In Fig.~\ref{fig:scan} we present 
a parameter scan over a wide range of $\beta$ values
from 
$\beta=0.1$ to $1$, and we vary the normalization parameter
$\mu$. We select a range of $\chi^2$ values in fair agreement
with the data (within $15\%$ of the best $\chi^2$ of the
$\Lambda$CDM model). All model parameters outside this range
are color-coded white. We note a a few areas of interest, 
all with values in the range between $\beta=0.1$ and $0.3$.
The point with a black square at $\beta=0.17$ is the model
from figure \ref{fig:sn}.

We are leaving the SN1a absolute magnitude as a free parameter in the analysis. 
There is a clear
valley at small values of $\mu$, covering $\beta$
from approximately 0.1 to 0.6.
Interestingly, some of the 
models have a slightly different evolution 
of the expansion
from the
standard $\Lambda$CDM model, both at high and low redshift, 
and we expect to quantify 
to which degree these models can be rejected with other
astronomical observations in a future paper.
Besides this valley, 
there are also a few models
at higher $\mu$
that fit the SN data fairly well, however, all with absolute
magnitudes  significantly different from  
the result from $\Lambda$CDM model of $-19.3$ 
and from observations
\cite{1993ApJ...413L.105P,1999AJ....118.2675R}. 
These models are indicated in figure 2 with a plus-sign
if $M > -19.0$ (high-$\mu$ region), and a minus-sign if $M<-19.5$ (below the valley).
\color{black}
If we instead would be using a
value of $n=-2.3$  we would
find a best fit value of $\beta = 0.27$
with parameters in fair agreement with the
SN1a data within the range
$0.2 < \beta < 0.4$. The most 
extreme (and most likely physically non-relevant)
possibility is the one of the 
undeveloped initial spectrum
of $n=+1$, which leads to a 
surprisingly good fit to the
expansion history with a fitted
values of $\beta = 1.4$ (with
reasonable values in the range
$1.2 < \beta < 2.0$).

\color{black}

\section{Discussion}

Our phenomenological description covers a wide range of underlying
models through the free parameter $\beta$. 
One concrete example is the assumption that the
changed energy is proportional to the surface area of
the black holes, and thus that the accelerated expansion
is driven by the growing black holes. 
It has been observed that there is a power-law relation between the BH mass and the velocity dispersion in the halo
\cite{2000ApJ...539L...9F, 2000ApJ...539L..13G}
\begin{equation}
    M_{\rm BH} \sim \sigma^{5.1}_{\rm halo}.
    \label{eq:Msigma}
    \end{equation}
\color{black}
Even though this relation is best established in the range 
$10^6 M_\odot < M < 10^{10} M_\odot$
we here extrapolate this relation to
all masses. We are thus not addressing
the physical mechanism establishing 
the connection between the galaxy masses
and the BH masses (which may be
energy feedback from supermassive BH during the galaxy formation process),
but we are instead merely taking this
as an observational fact. Also the 
significantly large number of stellar
sized BHs should change the 
details in the connection between 
the BH and galaxy masses beyond
eq.~(\ref{eq:Msigma}). In principle
one could improve on this simplification, 
however, we will  not  attempt this here.
\color{black}

Observations show that halo-mass and velocity dispersion are approximately connected through \cite{2022A&A...659A.126A}
\begin{equation}
    \frac{M_{\rm halo}}{10^{12} M_\odot} \approx \left( \frac{\sigma_{\rm halo}}{100 {\rm km/sec}} \right)^3    \, .
\end{equation}
Since the BH area is proportional to the BH mass squared, we thus
get $\beta \approx 2.4$.
If instead the relevant parameter is proportional to 
the BH area to the power $3/4$
\cite{1999PhRvL..82.4971C, 2021PhRvD.104g6024B}
then one should expect $\beta \approx 1.55$. From our analysis we
\color{black}
instead find $\beta \approx 0.2$, 
which is significantly smaller
than the BH prediction.
\color{black}
One should keep in mind that
there are large uncertainties here: The connection between
BH and halo masses has a spread, and also the connection
between the halo mass and velocity dispersion has
a non-trivial spread. 

\color{black}
Another model suggests a connection between
the accelerated expansion of the Universe and
the velocity dispersion of dark matter in cosmological haloes \cite{2021ApJ...910...98L, 2021MNRAS.506L..16H}, which
predicts $\beta = 0.5$. Since the mass-function can be
described with a scale-dependent power-spectrum with
a spectral index going from approximately $n \approx -2.5$ at
the smallest scales to $n\approx -1$ at galaxy cluster
scales 
\cite{2008cosm.book.....W, 2008gady.book.....B}
the true evolution is found by integrating over the
full mass distribution, rather than simplifying with
a single spectral index as we have done here.
We note that using a spectral index around $n=-2$
would lead to an accelerated expansion of the Universe
in fair agreement with SN1a data using $\beta =0.5$,
and we therefore conclude that the present analysis
cannot exclude the suggestion of 
refs.~\cite{2021ApJ...910...98L, 2021MNRAS.506L..16H}.
\color{black}

We have seen above that with an appropriate choice of the
free parameter $\beta$ one can get an expansion history
of the $\delta$CDM model 
in
fair agreement with that predicted in the standard 
$\Lambda$CDM model. 
This implies that all the observations mentioned in 
the introduction, including the CMB observations,
the growth of large scale structure, 
integrated Sachs-Wolfe effect etc, 
are in
agreement with predictions in this model. For instance,
if the CMB data is analysed with a $\Lambda CDM$ model,
then the result is that $\Omega_M \approx 0.3$ and $\Omega_\Lambda\approx 0.7$, and if the CMB instead is analysed with our 
model, then it will show that $\Omega_M \approx 0.3$
and that the accelerated expansion of the Universe
results from $\beta \approx 0.17$.
\color{black}
The fact that the expansion history
of the Universe in the 
standard
$\Lambda CDM$ model
to first approximation is
indistinguishable from that of the
$\delta CDM$ model with $\beta = 0.17$
also implies that the halo mass function
is essentially indentical in those two
models.
A related discussion on the growth of perturbations (in the linear regime)
was made by
\cite{2003MNRAS.346..573L} using
different description of the
general expansion history (see
also \cite{1998ApJ...508..483W}).
\color{black}

The main point of this paper is 
to demonstrate that one can get
an expansion history is fair agreement
with that of $\Lambda CDM$ model, 
entirely without using a cosmological
constant. This is exemplified by
plotting the full apparent magnitude
in Figure 1. Indeed the new model
first has deceleration at high
redshift (when there is very little
substructure), which then transitions
to an accelerated expansion in
the later Universe, just like the
$\Lambda CDM$ model.
Naturally, one should expect some
level of variation between the $\delta$CDM and $\Lambda$CDM models,
and it will be interesting in the future to investigate if
such differences may support the observational indications
that possibly
not even dynamic versions of the cosmological constant provide a self-consistent explanation of all the available cosmological data
\cite{2021ApJ...912..150D, 2021PhRvD.104h3519T, 2021JCAP...12..028K}.

A recent study of the evolution of BH masses has also
suggested a link between the BH mass increase and the
expansion of the Universe~\cite{2023ApJ...944L..31F}. 
That paper considered non-standard, singularity-free BHs, 
where stress-energy within these BH evolve with the expanding
Universe in such a way that the BH mass changes as
$M_{\rm BH} \sim a^3$, independent of the accretion
and merging of the galaxies. This description is very 
different from the one presented here (we consider
the evolution of structures to follow the accretion
and merging in the expanding Universe). However, it may
be possible to link our study to the one of 
\cite{2023ApJ...944L..31F} by not using the 
standard link between BH and haloes (as we use here) 
$M_{\rm BH} \sim M_{\rm halo}^5$. We will leave
such detailed comparisons for a future study.

\color{black}
Several limitations of the present approach relate to the calculation of
the distribution of the small scale
structure. First of all, whereas the
Press-Schechter formalism was the
first and simplest method to analytically calculate
the mass-function, it has been 
demonstrated, in particular through
the use of numerical 
cosmological simulations, 
that both the mass-dependence and
redshift evolution has somewhat different
properties than those predicted by 
PS \cite{2002MNRAS.329...61S, 2008ApJ...688..709T, 2021ApJ...922...89S}.
Secondly, whereas we here simplify the
full mass function as a simple power-law,
in reality one should integrate over the
full distribution function. 

\color{black}

In this discussion it was  assumed that all BHs follow
the standard correlation with 
$M_{\rm BH} \sim \sigma^{5.1}_{\rm halo}$. It may
be that the early Universe contains BHs 
significantly more massive
\cite{2013MNRAS.432.3438A, 2019MNRAS.486.2336D, 2022MNRAS.511.3751H, 2023ApJ...955L..24G}, 
which in particular may change the details of the 
evolution of the Universe.

\section{Conclusions}
In order to 
ensure that effective field theories remain consistent,
a relationship between the UV cut-off and the IR physics
has been proposed \cite{1999PhRvL..82.4971C}, which 
suggests a relationship between Dark Energy and Black Holes.
In order to test this connection we
present a time-dependent calculation, which includes
the formation and evolution of all
structure formation (which links to the evolving
masses of BHs) in the expanding Universe. 
By comparison with cosmological SN1a data, we 
\color{black}
find that the simplest models 
of \cite{1999PhRvL..82.4971C, 2021PhRvD.104g6024B},
where we
extrapolated the observed BH-galaxy masses to be valid at all masses, are not
in agreement with the expansion history
as measured throught SN1a.
Instead we find that another simple
model where the energy term is
$\Delta E \sim M ^{\beta+1}$
is in fairly good agreement with the SN1a
data using $\beta \approx 0.2$.
The limitations of the description presented above, 
which are
dominated by the assumption that the mass-spectrum
of haloes can be simplified by a single spectral
index, 
implies that
we cannot exclude the possibility that the
accelerated expansion may be driven by an effect
driven by the velocity dispersions of 
galaxies
\cite{2021ApJ...910...98L, 2021MNRAS.506L..16H}.
\color{black}

\acknowledgments{It is a pleasure thanking Zhen Li for interesting discussions in the
early phase of this project. We thank the anonymous 
referees for excellent suggestions which improved the paper.}

\conflictsofinterest{The authors declare no conflicts of interest.}

\reftitle{References}

\PublishersNote{}

\begin{thebibliography}{999}


\bibitem[Perlmutter et al.(1999)]{1999ApJ...517..565P} Perlmutter, S., Aldering, G., Goldhaber, G., et al.\,
 “Measurements of $\Omega$ and $\Lambda$ from 42 High-Redshift Supernovae”, The Astrophysical Journal,  {\bf 1999}, vol. 517, no. 2, pp. 565–586



\bibitem[Riess et al.(1998)]{1998AJ....116.1009R} Riess, A.~G., Filippenko, A.~V., Challis, P., et al.\ 
“Observational Evidence from Supernovae for an Accelerating Universe and a Cosmological Constant”, The Astronomical Journal, {\bf 1998},
vol. 116, no. 3, pp. 1009–1038

\bibitem[Komatsu et al.(2011)]{2011ApJS..192...18K} Komatsu, E., Smith, K.~M., Dunkley, J., et al.\ “Seven-year Wilkinson Microwave Anisotropy Probe (WMAP) Observations: Cosmological Interpretation”, The Astrophysical Journal Supplement Series, {\bf  2011} vol. 192, no. 2, 2011. doi:10.1088/0067-0049/192/2/18.


\bibitem[Percival et al.(2010)]{2010MNRAS.401.2148P} Percival, W.~J., Reid, B.~A., Eisenstein, D.~J., et al.\ “Baryon acoustic oscillations in the Sloan Digital Sky Survey Data Release 7 galaxy sample”, Monthly Notices of the Royal Astronomical Society, {\bf  2010}, vol. 401, no. 4, pp. 2148–2168, 2010. doi:10.1111/j.1365-2966.2009.15812.x.


\bibitem[Blake et al.(2011)]{2011MNRAS.415.2876B} Blake, C., Brough, S., Colless, M., et al.\ The WiggleZ Dark Energy Survey: the growth rate of cosmic structure since redshift z=0.9”, Monthly Notices of the Royal Astronomical Society, {\bf  2011}, vol. 415, no. 3, pp. 2876–2891. doi:10.1111/j.1365-2966.2011.18903.x.


\bibitem[Larson et al.(2011)]{2011ApJS..192...16L} Larson, D., Dunkley, J., Hinshaw, G., et al.\ 
“Seven-year Wilkinson Microwave Anisotropy Probe (WMAP) Observations: Power Spectra and WMAP-derived Parameters”, The Astrophysical Journal Supplement Series, {\bf  2011}, vol. 192, no. 2, 2011. doi:10.1088/0067-0049/192/2/16.


\bibitem[Hicken et al.(2009)]{2009ApJ...700.1097H} Hicken, M., Wood-Vasey, W.~M., Blondin, S., et al.\ 
“Improved Dark Energy Constraints from ~100 New CfA Supernova Type Ia Light Curves”, The Astrophysical Journal, {\bf  2009}, vol. 700, no. 2, pp. 1097–1140, 2009. doi:10.1088/0004-637X/700/2/1097.



\bibitem[Blake et al.(2011)]{2011MNRAS.418.1725B} Blake, C., Glazebrook, K., Davis, T.~M., et al.\ 
“The WiggleZ Dark Energy Survey: measuring the cosmic expansion history using the Alcock-Paczynski test and distant supernovae”, Monthly Notices of the Royal Astronomical Society, {\bf  2011}, vol. 418, no. 3, pp. 1725–1735, 2011. doi:10.1111/j.1365-2966.2011.19606.x.



\bibitem[Weinberg(1989)]{1989RvMP...61....1W} Weinberg, S.\ 
“The cosmological constant problem”, Reviews of Modern Physics, {\bf 1989 }, vol. 61, no. 1, pp. 1–23. doi:10.1103/RevModPhys.61.1.





\bibitem[Peebles \& Ratra(2003)]{2003RvMP...75..559P} Peebles, P.~J. \& Ratra, B.\ 
“The cosmological constant and dark energy”, Reviews of Modern Physics, {\bf  2003}, vol. 75, no. 2, pp. 559–606. doi:10.1103/RevModPhys.75.559.


\bibitem[Giudice(2017)]{2017arXiv171007663G} Giudice, G.~F.\ 
“The Dawn of the Post-Naturalness Era”, {\bf  2017}. doi:10.48550/arXiv.1710.07663.


\bibitem[Cohen et al.(1999)]{1999PhRvL..82.4971C} Cohen, A.~G., Kaplan, D.~B., \& Nelson, A.~E.\ 
“Effective Field Theory, Black Holes, and the Cosmological Constant”, Physical Review Letters, {\bf  1999 }, vol. 82, no. 25, pp. 4971–4974. doi:10.1103/PhysRevLett.82.4971.


\bibitem[Bekenstein(1973)]{1973PhRvD...7.2333B} Bekenstein, J.~D.\ 
“Black Holes and Entropy”, Physical Review D, {\bf  1973}, vol. 7, no. 8, pp. 2333–2346. doi:10.1103/PhysRevD.7.2333.


\bibitem[Bekenstein(1981)]{1981PhRvD..23..287B} Bekenstein, J.~D.\ 
“Universal upper bound on the entropy-to-energy ratio for bounded systems”, Physical Review D, {\bf 1981 }, vol. 23, no. 2, pp. 287–298, . doi:10.1103/PhysRevD.23.287.



\bibitem[Blinov \& Draper(2021)]{2021PhRvD.104g6024B} Blinov, N. \& Draper, P.\ 
“Densities of states and the Cohen-Kaplan-Nelson bound”, Physical Review D, {\bf  2021}, vol. 104, no. 7. doi:10.1103/PhysRevD.104.076024.


\bibitem[Abel \& Dienes(2021)]{2021PhRvD.104l6032A} Abel, S. \& Dienes, K.~R.\ 
“Calculating the Higgs mass in string theory”, Physical Review D, {\bf  2021}, vol. 104, no. 12. doi:10.1103/PhysRevD.104.126032.


\bibitem[Craig \& Koren(2020)]{2020JHEP...03..037C} Craig, N. \& Koren, S.\ 
“IR dynamics from UV divergences: UV/IR mixing, NCFT, and the hierarchy problem”, Journal of High Energy Physics, {\bf  2020}, vol. 2020, no. 3, . doi:10.1007/JHEP03(2020)037.


\bibitem[Loeve et al.(2021)]{2021ApJ...910...98L} Loeve, K., Nielsen, K.~S., \& Hansen, S.~H.\ 
“Consistency Analysis of a Dark Matter Velocity-dependent Force as an Alternative to the Cosmological Constant”, The Astrophysical Journal, {\bf  2021}, vol. 910, no. 2. doi:10.3847/1538-4357/abe5a2.


\bibitem[Hansen(2021)]{2021MNRAS.506L..16H} Hansen, S.~H.\ 
“A force proportional to velocity squared derived from spacetime algebra”, Monthly Notices of the Royal Astronomical Society, {\bf  2021}, vol. 506, no. 1, pp. L16–L19. doi:10.1093/mnrasl/slab065.


\bibitem[Press \& Schechter(1974)]{1974ApJ...187..425P} Press, W.~H. \& Schechter, P.\ 
“Formation of Galaxies and Clusters of Galaxies by Self-Similar Gravitational Condensation”, The Astrophysical Journal, {\bf  1974}, vol. 187, pp. 425–438. doi:10.1086/152650.
´

\bibitem[Linder \& Jenkins(2003)]{2003MNRAS.346..573L} Linder, E.~V. \& Jenkins, A.\ 2003, MNRAS, 346, 573. doi:10.1046/j.1365-2966.2003.07112.x


\bibitem[Weinberg(2008)]{2008cosm.book.....W} Weinberg, S.\ 2008, Cosmology, by Steven Weinberg. ISBN 978-0-19-852682-7. Published by Oxford University Press, Oxford, UK, 2008.

\bibitem[Binney \& Tremaine(2008)]{2008gady.book.....B} Binney, J. \& Tremaine, S.\ 2008, Galactic Dynamics: Second Edition, by James Binney and Scott Tremaine. ISBN 978-0-691-13026-2 (HB). Published by Princeton University Press, Princeton, NJ USA, 2008.


\bibitem[Heath(1977)]{1977MNRAS.179..351H} Heath, D.~J.\ 
“The growth of density perturbations in zero pressure Friedmann-Lemaître universes.”, Monthly Notices of the Royal Astronomical Society, {\bf  1977}, vol. 179, pp. 351–358. doi:10.1093/mnras/179.3.351.


\bibitem[Carroll et al.(1992)]{1992ARA&A..30..499C} Carroll, S.~M., Press, W.~H., \& Turner, E.~L.\ 
 “The cosmological constant.”, Annual Review of Astronomy and Astrophysics, {\bf  1992 }, vol. 30, pp. 499–542. doi:10.1146/annurev.aa.30.090192.002435.



\bibitem[Longair(2008)]{2008gafo.book.....L} Longair, M.~S.\ {\bf  2008}, Galaxy Formation, Berlin: Springer.  ISBN 978-3-540-73477-2


\bibitem[Riess et al.(2004)]{2004ApJ...607..665R} Riess, A.~G., Strolger, L.-G., Tonry, J., et al.\ 2004, ApJ, 607, 665. doi:10.1086/383612



\bibitem[Scolnic et al.(2022)]{2022ApJ...938..113S} Scolnic, D., Brout, D., Carr, A., et al.\ 
“The Pantheon+ Analysis: The Full Data Set and Light-curve Release”, The Astrophysical Journal, {\bf  2022}, vol. 938, no. 2. doi:10.3847/1538-4357/ac8b7a.


\bibitem[Phillips(1993)]{1993ApJ...413L.105P} Phillips, M.~M.\ 
“The Absolute Magnitudes of Type IA Supernovae”, The Astrophysical Journal, {\bf  1993}, vol. 413, p. L105. doi:10.1086/186970.


\bibitem[Riess et al.(1999)]{1999AJ....118.2675R} Riess, A.~G., Filippenko, A.~V., Li, W., et al.\ 
“The Rise Time of Nearby Type IA Supernovae”, The Astronomical Journal, {\bf  1999}, vol. 118, no. 6, pp. 2675–2688. doi:10.1086/301143.


\bibitem[Ferrarese \& Merritt(2000)]{2000ApJ...539L...9F} Ferrarese, L. \& Merritt, D.\ “A Fundamental Relation between Supermassive Black Holes and Their Host Galaxies”, The Astrophysical Journal, {\bf  2000}, vol. 539, no. 1, pp. L9–L12. doi:10.1086/312838.


\bibitem[Gebhardt et al.(2000)]{2000ApJ...539L..13G} Gebhardt, K., Bender, R., Bower, G., et al.\ “A Relationship between Nuclear Black Hole Mass and Galaxy Velocity Dispersion”, The Astrophysical Journal, {\bf  2000},  vol. 539, no. 1, pp. L13–L16. doi:10.1086/312840.


\bibitem[Aguado-Barahona et al.(2022)]{2022A&A...659A.126A} Aguado-Barahona, A., Rubi{\~n}o-Mart{\'\i}n, J.~A., Ferragamo, A., et al.\ “Velocity dispersion and dynamical masses for 388 galaxy clusters and groups. 
Calibrating the $M_{SZ}$ - $M_{dyn}$ scaling relation for the PSZ2 sample”, Astronomy and Astrophysics, {\bf  2022}, vol. 659. doi:10.1051/0004-6361/202039980.



\bibitem[Wang \& Steinhardt(1998)]{1998ApJ...508..483W} Wang, L. \& Steinhardt, P.~J.\ 1998, ApJ, 508, 483. doi:10.1086/306436



\bibitem[Dainotti et al.(2021)]{2021ApJ...912..150D} Dainotti, M.~G., De Simone, B., Schiavone, T., et al.\ 
“On the Hubble Constant Tension in the SNe Ia Pantheon Sample”, The Astrophysical Journal, {\bf  2021}, vol. 912, no. 2. doi:10.3847/1538-4357/abeb73.


\bibitem[Teng et al.(2021)]{2021PhRvD.104h3519T} Teng, Y.-P., Lee, W., \& Ng, K.-W.\ 
“Constraining the dark-energy equation of state with cosmological data”, Physical Review D, {\bf  2021}, vol. 104, no. 8. doi:10.1103/PhysRevD.104.083519.


\bibitem[Krolewski et al.(2021)]{2021JCAP...12..028K} Krolewski, A., Ferraro, S., \& White, M.\ 
“Cosmological constraints from unWISE and Planck CMB lensing tomography”, Journal of Cosmology and Astroparticle Physics, {\bf  2021}, vol. 2021, no. 12. doi:10.1088/1475-7516/2021/12/028.


\bibitem[Farrah et al.(2023)]{2023ApJ...944L..31F} Farrah, D., Croker, K.~S., Zevin, M., et al.\ 
“Observational Evidence for Cosmological Coupling of Black Holes and its Implications for an Astrophysical Source of Dark Energy”, The Astrophysical Journal, {\bf  2023}, vol. 944, no. 2. doi:10.3847/2041-8213/acb704.




\bibitem[Sheth \& Tormen(2002)]{2002MNRAS.329...61S} Sheth, R.~K. \& Tormen, G.\ 2002, MNRAS, 329, 61. doi:10.1046/j.1365-8711.2002.04950.x

\bibitem[Tinker et al.(2008)]{2008ApJ...688..709T} Tinker, J., Kravtsov, A.~V., Klypin, A., et al.\ 2008, ApJ, 688, 709. doi:10.1086/591439

\bibitem[Shirasaki et al.(2021)]{2021ApJ...922...89S} Shirasaki, M., Ishiyama, T., \& Ando, S.\ 2021, ApJ, 922, 89. doi:10.3847/1538-4357/ac214b







\bibitem[Agarwal et al.(2013)]{2013MNRAS.432.3438A} Agarwal, B., Davis, A.~J., Khochfar, S., et al.\ “Unravelling obese black holes in the first galaxies”, Monthly Notices of the Royal Astronomical Society, {\bf  2013}, vol. 432, no. 4, pp. 3438–3444. doi:10.1093/mnras/stt696.





\bibitem[Dayal et al.(2019)]{2019MNRAS.486.2336D} Dayal, P., Rossi, E.~M., Shiralilou, B., et al.\ “The hierarchical assembly of galaxies and black holes in the first billion years: predictions for the era of gravitational wave astronomy”, Monthly Notices of the Royal Astronomical Society, {\bf  2019}, vol. 486, no. 2, pp. 2336–2350. doi:10.1093/mnras/stz897.



\bibitem[Habouzit et al.(2022)]{2022MNRAS.511.3751H} Habouzit, M., Onoue, M., Ba{\~n}ados, E., et al.\ “Co-evolution of massive black holes and their host galaxies at high redshift: discrepancies from six cosmological simulations and the key role of JWST”, Monthly Notices of the Royal Astronomical Society, {\bf  2022}, vol. 511, no. 3, pp. 3751–3767. doi:10.1093/mnras/stac225.


\bibitem[Goulding et al.(2023)]{2023ApJ...955L..24G} Goulding, A.~D., Greene, J.~E., Setton, D.~J., et al.\ “UNCOVER: The Growth of the First Massive Black Holes from JWST/NIRSpec-Spectroscopic Redshift Confirmation of an X-Ray Luminous AGN at z = 10.1”, The Astrophysical Journal, {\bf  2023}, vol. 955, no. 1. doi:10.3847/2041-8213/acf7c5.


\end{thebibliography}
\end{document}